# Ambient pressure growth of bilayer nickelate single crystals with superconductivity over 90 K under high pressure

Feiyu Li,[1†] Zhenfang Xing,[2†] Di Peng,[3,2*] Jie Dou,[4,5] Ning Guo,[6] Liang Ma,[4,7,8] Yulin Zhang,[1] Lingzhen Wang,[1] Jun Luo,[4] Jie Yang,[4] Jian Zhang,[1] Tieyan Chang,[9] Yu-Sheng Chen,[9] Weizhao Cai,[10,11] Jinguang Cheng,[4,5] Yuzhu Wang,[12] Yuxin Liu,[2] Tao Luo,[2] Naohisa Hirao,[13] Takahiro Matsuoka,[14] Hirokazu Kadobayashi,[13] Zhidan Zeng,[2] Qiang Zheng,[6] Rui Zhou,[4,5] Qiaoshi Zeng,[2,3*] Xutang Tao,[1*] and Junjie Zhang[1*]

[1]State Key Laboratory of Crystal Materials, Institute of Crystal Materials, Shandong University, Jinan, Shandong 250100, China

[2]Center for High Pressure Science and Technology Advanced Research, Shanghai 201203, China

[3]Shanghai Key Laboratory of Material Frontiers Research in Extreme Environments (MFree), Shanghai Advanced Research in Physical Sciences (SHARPS), Shanghai 201203, China

[4]Beijing National Laboratory for Condensed Matter Physics and Institute of Physics, Chinese Academy of Sciences, Beijing 100190, China

[5]School of Physical Sciences, University of Chinese Academy of Sciences, Beijing 100190, China

[6]CAS Key Laboratory of Standardization and Measurement for Nanotechnology, National Center for Nanoscience and Technology, Beijing 100190, China

[7]Key Laboratory of Materials Physics, Ministry of Education, School of Physics, Zhengzhou University, Zhengzhou 450001, China

[8]Institute of Quantum Materials and Physics, Henan Academy of Sciences, Zhengzhou 450046, China

[9]NSF's ChemMatCARS, The University of Chicago, Lemont, Illinois 60439, United States

[10]School of Materials and Energy, University of Electronic Science and Technology of China, Chengdu 611731, Sichuan, China

[11]Huzhou Key Laboratory of Smart and Clean Energy, Yangtze Delta Region Institute (Huzhou), University of Electronic Science and Technology of China, Huzhou 313001, China

[12]Shanghai Synchrotron Radiation Facility, Shanghai Advanced Research Institute, Chinese Academy of Sciences, Shanghai 201204, China

[13]Japan Synchrotron Radiation Research Institute, Sayo, Hyogo 679-5198, Japan

[14]NanoTerasu Promotion Division, Japan Synchrotron Radiation Research Institute (JASRI), Sendai, Miyagi 980-8572, Japan

[†]These authors contributed equally: Feiyu Li and Zhenfang Xing

[*]Correspondence to: di.peng@hpstar.ac.cn, zengqs@hpstar.ac.cn, txt@sdu.edu.cn and junjie@sdu.edu.cn

**Abstract:** Recently, the Ruddlesden-Popper bilayer nickelate $La_3Ni_2O_7$ has been discovered as a high temperature superconductor with $T_c$ near 80 K above 14 GPa.[1-3] The search for nickelate superconductors with higher $T_c$, the preparation of high-quality single crystals, and the removal of high-pressure conditions including single crystal growth under high gas pressure and achievement of high $T_c$ superconductivity under high pressure, are the most challenging tasks. Here, we present ambient pressure flux growth of bilayer nickelate single crystals with superconductivity up to 91 K under high pressure. Single crystals of bilayer $La_{3-x}R_xNi_2O_{7-\delta}$ (R= Pr-Er; $x \leq 2.7$) with dimensions up to 220 μm on the edge were successfully grown using flux method at atmosphere conditions. Single crystal X-ray diffraction, nuclear quadrupole resonance, energy dispersion spectroscopy and scanning transmission electron microscopy measurements evidenced high quality of bilayer $La_2SmNi_2O_{7-\delta}$ single crystals in both average structure and local structure. Superconductivity has been observed in high pressure resistivity measurements of annealed $La_2SmNi_2O_{7-\delta}$ single crystals with $T_c^{onset}$ up to 91 K, which is the highest among the known superconducting nickelates. Our results not only demonstrate a new and easy-to-access method for synthesizing high-quality bilayer nickelate single crystals, but also providing a direction for discovering superconducting nickelates with higher $T_c$.

**Main text**

Understanding the mechanism of high temperature superconductivity and discovery of high $T_c$ superconductors in transition metal oxides beyond cuprates are among the frontiers in the fields of condensed matter physics and materials science.[4-6] As a neighbor of copper in the periodic table, nickelates have been long sought for high-$T_c$ superconductivity.[7-9] The square planar trilayer nickelate $Pr_4Ni_3O_8$, which exhibits quasi-two-dimensional square lattice, low spin, large orbital polarization and strange metal behavior, was reported to be a close analogue of the over-doped cuprates.[9] A breakthrough was achieved in 2019 by Li et al., who reported 9-15 K superconductivity in the thin films of infinite-layer $Nd_{0.8}Sr_{0.2}NiO_2$,[10] initiating the "Nickel Age" of Superconductivity.[11-17] Unfortunately, no superconductivity in bulk samples of infinite-layer nickelates has been reported after more than five-year intense research.[18-20] The highest $T_c$ in square planar layered nickelates reported up to date is $Sm_{1-x-y-z}Eu_xCa_ySr_zNiO_2$ with $T_c^{onset}$=37 K,[21] which is still far below the boiling point of liquid nitrogen (77 K).

Recently, signature of superconductivity near 80 K was reported by Sun et al. in bilayer Ruddlesden-Popper (R-P) $La_3Ni_2O_7$ single crystals under a pressure of 14.0-43.5 GPa.[1] This discovery triggered tremendous interest in nickelates in both experimental and theoretical aspects.[17,22-30] Zero resistance and Meissner effect, two hallmarks of superconductivity, were confirmed under high pressure in bilayer $La_3Ni_2O_7$ single crystals[2,3] and $La_2PrNi_2O_7$ polycrystalline powders.[31] Substitutions of Pr for La in $La_3Ni_2O_7$ effectively improve sample quality of polycrystalline powders by inhibiting the intergrowth of different R-P phases found in $La_3Ni_2O_7$, resulting in nearly pure bilayer $La_2PrNi_2O_7$ phase with $T_c^{zero}$=60 K and $T_c^{onset}$=82 K at 18-20 GPa.[31] More recently, by mimicking high pressure conditions, ultrathin films of $La_3Ni_2O_7$ have been reported to exhibit superconductivity at ambient pressure.[26,30] However, whether nickelate high-$T_c$ superconductivity can be achieved in bulk samples – polycrystalline powders and single crystals – at ambient pressure remains an open question.

Another important fundamental question is how to further increase $T_c$ in nickelate superconductors. Up to date, the highest $T_c$ was reported to be 83 K for $La_3Ni_2O_7$ single crystals[32] and 86 K for $La_3Ni_2O_7$ powders[33], and 82 K for $La_2PrNi_2O_7$ powders.[31] It has been proposed to search for higher $T_c$ in trilayer nickelates by analogy with cuprates, i.e., the $T_c$ of bilayer cuprates is lower than trilayer cuprates, which have the highest $T_c$ among all cuprates.[34] Unexpectedly, the $T_c^{onset}$ of trilayer $R_4Ni_3O_{10}$ (R=La and Pr) were reported to be 25-39 K,[27,35,36] much lower than the bilayer $La_3Ni_2O_7$.[1,37] The different layer dependence of superconductivity between nickelates and cuprates remains an open question.

We propose to explore superconducting nickelates with higher $T_c$ by introducing chemical pressure via cation substitution using smaller rare earth elements in bilayer $La_3Ni_2O_7$. Employing chemical pressure to increase $T_c$ has been realized in iron-based superconductors, i.e., the optimal $T_c$ increases with the decreasing radii of the rare-earth metal ions, reaching the highest $T_c$=55 K in the doped SmFeAsO system.[38] In nickelate systems, insulator-to-metal transition induced by chemical pressure have been reported in the square planar trilayer system by Zhang et al., proving the capability of tuning electronic structure using chemical pressure.[9] Theoretical calculations on bilayer $R_3Ni_2O_7$ (R=La-Sm) predicted different trends on $T_c$.[39,40] Zhang et al. reported that $La_3Ni_2O_7$ is already the "optimal" candidate, and $T_c$ decreases as the radius of rare-earth (RE) ions decreases.[39] In sharp contrast, Pan et al. predicted that the $T_c$ increases from La to Sm and a nearly doubled $T_c$ can be achieved for $Sm_3Ni_2O_7$.[40] Experimentally, no $R_3Ni_2O_7$ (R=La-Sm) except R=La have been synthesized.[31,41] With these in mind, we focused our efforts on optimizing and investigating $La_{3-x}R_xNi_2O_{7-\delta}$ (R=Pr-Er, $x \leq 2.7$) single crystals, which we present in detail here.

**Single crystal growth of bilayer nickelates at ambient pressure**

We report a new and easy-to-access ambient-pressure growth method to synthesize bilayer nickelate single crystals. Previously, $La_3Ni_2O_7$ single crystals were grown under 10-18 bar of oxygen pressure using floating zone techniques.[41,42] Sample issues including inhomogeneity, impurities, intergrowth and oxygen vacancies[2,43] have been reported. What's worse, the hybrid $La_2NiO_4{\cdot}La_4Ni_3O_{10}$ competes during the floating-zone growth of bilayer $La_3Ni_2O_7$, making it challenging to obtain a pure phase.[44-46] Following our previous recipe on the growth of trilayer $La_4Ni_3O_{10}$[47] and hybrid $La_2NiO_4{\cdot}La_3Ni_2O_7$,[48] we succeeded in growing bilayer $La_3Ni_2O_7$ single crystals at ambient pressure (see **Methods and Table I**). **Fig.1a** shows a scheme of the growth setup via evaporation of flux. **Fig.1b** shows an SEM image of a typical single crystal of $La_3Ni_2O_7$ with dimensions of 120 μm on the edge. The as-grown $La_3Ni_2O_{7-\delta}$ single crystals belong to the monoclinic $P2_1/m$ space group (**Fig.1d and Table S1**), which is lower than previously reported *Amam*[1] (**Fig. S1**) likely due to free growth from solution and/or oxygen deficiency. However, the out-of-plane Ni-O-Ni bond angle is 168.5(3)° (**Fig.1e**), the same as that in *Amam*.[1] Rietveld refinement on X-ray powder diffraction data collected at room temperature of pulverized as-grown single crystals, as shown in **Fig.1h**, verified the single crystal structural model. Compared with high pressure floating zone growth, [44-46] we completely remove the competing phase of $La_2NiO_4{\cdot}La_4Ni_3O_{10}$, and signicantly improve sample quality (**Fig.S2 and Fig. 3d**); however, there still exist two issues: (i) hybrid $La_2NiO_4{\cdot}La_3Ni_2O_7$[48] single crystals appear as a secondary phase (**Fig.1h**), and (ii) intergrowth of R-P phases is clearly seen from nuclear quadrupole resonance (NQR) measurements (**Fig.3d**).

We then explored if substitution of La using smaller rare earth elements can inhibit intergrowth of R-P phases in single crystal growth, like the case of $La_2PrNi_2O_7$ polycrystalline powders reported by Wang et al.[31] We started with La:R=2:1 (R=Pr-Er) in flux growth. Black and shiny single crystals of $La_{3-x}R_xNi_2O_7$ (R=Pr-Er) were obtained by removing remaining flux after growth (**Fig.1c** and **Fig. S3**). The value of $x$ in $La_{3-x}R_xNi_2O_{7-\delta}$ (R=Pr-Er) was determined by energy dispersive spectroscopy (EDS) (**Fig.S3**) to be $x<1$ for R=Eu-Er and $x\sim1$ for R = Pr-Sm, implying that the amounts of substitutions in the La sites can be larger as the size of rare earth is close to La. Interestingly, pure phase was obtained for the Sm case (**Fig.1i**). By suppressing the growth of hybrid R-P phase (**Fig.S4 and Fig.S5**), the size of single crystals of $La_2SmNi_2O_7$ was increased to 220 μm on the edge (**Fig.1c**), almost twice compared with $La_3Ni_2O_7$. **Fig.1f** shows the crystal structure of $La_2SmNi_2O_{7-\delta}$ obtained from single crystal X-ray diffraction, which also has monoclinic structure $P2_1/m$ (**Table S1**). Instead of random distribution, the substitutions of Sm preferentially occupy the La sites between bilayers (**Fig.1f**). The out-of-plane Ni-O-Ni bond angle decreases to 164.2(5)° (**Fig.1g**), further deviating from 180°. The calculated bond valence sum (BVS) values of Ni are 2.67 and 2.70, similar to the calculated values of $La_3Ni_2O_7$ (2.67 and 2.69), indicating that the incorporation of Sm does not change the valence state of Ni and the valence of Sm is 3+. Rietveld refinement on powder X-ray diffraction data collected at room temperature of pulverized as-grown $La_2SmNi_2O_{7-\delta}$ single crystals (**Fig.1i**) converged to $R_{exp}$ = 3.27%, $R_{wp}$ = 7.11%, GOF = 2.18, corroborating the single crystal structural model.

We move to maximize the value of $x$ for R=Pr-Sm by optimizing growth conditions. Lebail fit was used to extract the lattice parameters of the pulverized as-grown single crystals. We found that the increasing of $a$ and decreasing of $b$, $c$ and $V$ correlate with the increasing of $x$ (**Fig. S6**). Through EDS, the maximum $x$ was determined to be 2.7, 2.1 and 1.4 for Pr, Nd and Sm, respectively (**Fig. S7**). **Table I** summarizes the growth result of all $La_{3-x}R_xNi_2O_{7-\delta}$ (R=La-Er) with floating zone growth of $La_3Ni_2O_7$ as a comparison. **Fig.2a** shows the maximum substitution for various rare earth elements obtained from EDS. Clearly, the maximum $x$ decreases with decreasing the size of rare earth ions; however, the molar ratio of (La+R) to Ni maintains 3:2. **Fig.2b-c** show the lattice parameters as a function of rare earth ions at the maximum $x$. **Fig.2d** shows the variation of averaged in-plane lattice constant of $La_{3-x}R_xNi_2O_{7-\delta}$ (R=La-Er) calculated by $a_{average} = (a^2 + b^2)^{1/2}/2$ along with the lattice constants in the superconducting states.[26] With the decreasing of rare earth ions, the in-plane lattice constant shows a non-monotonic trend, reaching a minimum at R=Nd with $x$=2.1. Compared with the superconducting state[26], there is still a big gap in the in-plane lattice constants, implying that it is difficult to observe superconductivity in the $La_{3-x}R_xNi_2O_{7-\delta}$ (R=La-Er) bulk crystals at ambient pressure, and high-pressure is needed to achieve superconductivity if exist.

### Crystal quality of $La_2SmNi_2O_{7-\delta}$ single crystals

Prior to high pressure measurements, we evaluate the crystal quality of $La_2SmNi_2O_{7-\delta}$ single crystals on both the average structure level and local structure level. **Fig.3a** shows the EDS mapping on a typical $La_2SmNi_2O_7$ single crystal with dimensions of ~200 μm on the edge. The compositions at different positions over the range of the whole crystal are identical, demonstrating high homogeneity of the distribution of La/Sm and Ni. **Fig.3b** and **Fig.3c** show the reconstructed (0*kl*) and (*h*0*l*) planes of a $La_2SmNi_2O_{7-\delta}$ single crystal measured by single crystal X-ray diffraction at 296 K. The observed peaks are neat and obey the selection rule of $P2_1/m$.

Next, we employed nuclear quadrupole resonance (NQR) as a sensitive, global probe to investigate possible intergrowth of R-P phases,[31,49] as shown in **Fig.3d**. For as-grown single-crystal samples of $La_3Ni_2O_7$, four distinct resonance lines were observed, consistent with previous NQR measurements on powder samples[31]. These resonance peaks, from low to high frequency, correspond to the $La^{4310}(2)$, $La^{327-i}(2)$, $La^{4310-i}(2)$, and $La^{327}(2)$ sites. Here, $La^{327-i}(2)$ and $La^{4310-i}(2)$ are related to the intergrowth between the $La_3Ni_2O_7$ and $La_4Ni_3O_{10}$ phases.[31] Notably, the NQR linewidth of the $La_3Ni_2O_7$ single crystal is more than an order of magnitude narrower than that of polycrystalline samples, indicating significantly improved crystallinity in our single-crystal samples (**Fig. S8**). For both as-grown and annealed $La_2SmNi_2O_{7-\delta}$, the NQR spectrum exhibits only a broad resonance peak, rather than the four distinct lines observed in $La_3Ni_2O_7$. This behavior is similar to that of $La_2PrNi_2O_7$ polycrystalline samples.[31] Our results also indicate that annealing process of $La_2SmNi_2O_{7-\delta}$, even with $P_{O2}$ =15 bar, does not introduce other R-P phases in the sample. A narrower NQR spectrum with a shift toward higher frequencies is observed in the annealed sample, which might be attributed to slightly increased oxygen content. Our single crystal X-ray diffraction analysis revealed that Sm atoms preferentially substitute for interlayer La atoms (La(2) sites), thereby introducing positional disorder at the La(2) sites. This increased disorder results in a large NQR linewidth as we observed in **Fig.3d**. Thus, our NQR measurements indicate significantly reduced intergrowth of R-P phases in $La_2SmNi_2O_{7-\delta}$ single crystals, similar to that in $La_2PrNi_2O_7$ polycrystalline powders,[31] compared with $La_3Ni_2O_7$.[31]

We further investigated the local structure of $La_2SmNi_2O_{7-\delta}$ single crystals using scanning transmission electron microscopy (STEM). A typical high-angle annular dark-field (HAADF)-STEM image in the projection [110] (**Fig.3e)** shows that $La_2SmNi_2O_{7-\delta}$ single crystals have perfectly ordered bilayer alternating stacks on the scale of tens of nanometers. To verify that it is the universal feature in such single crystals, more than twenty different regions in two single crystals were selected for HAADF imaging, all revealing perfectly ordered stacking sequences without intergrowth. The chemical distribution of La, Sm and Ni can be seen from the EDS diagram in **Fig.3f**. The results show that Sm preferentially occupies the La site between bilayers, consistent with the result of single crystal X-ray diffraction (**Fig.1f**).

### Superconductivity at 91 K of $La_2SmNi_2O_{7-\delta}$ under high pressure

EDS mapping, single crystal X-ray diffraction, NQR and real-space imaging via STEM evidence high quality of our bilayer $La_2SmNi_2O_{7-\delta}$ single crystals grown from ambient flux growth. The oxygen content is expected to be deficient, thus, we annealed the as-grown single crystals at an oxygen pressure of 1.5 bar for ten days to increase oxygen content. **Fig.4** presents evidence of pressure-induced superconductivity in annealed $La_2SmNi_2O_{7-\delta}$ single crystals with $T_c^{onset}$ up to 91 K, which is the highest among the known superconducting nickelates (**Fig S9**).[1,17,32,33] Temperature-dependent resistance of Crystal #1 under pressure in a paraffin-filled diamond anvil cell (DAC) is shown in **Fig.4a**. An apparent drop in resistance starting at 91 K is observed, indicating a superconducting transition. At 10 K, the residual resistance is only 5.5% of that prior to the transition. The observation of similar behaviors in the electrical resistance measured along two orthogonal directions separated by approximately a 90º angle demonstrates excellent homogeneity in the single crystal and excludes the possibility of filamentary superconductivity in our samples. Moreover, the metallic normal state of Crystal #1 and #2 (**Fig S10**) is also consistent with the well-characterized measurements in nickel-based superconductors.[1,2] The observation of similar superconducting transitions in various directions during the measurements of Crystal #3, as shown in **Fig S11**, indicates a consistent characteristic of the superconducting transition across different orientations. This further substantiates the bulk superconducting nature of the sample. Furthermore, the suppression of the superconducting transition by the applied magnetic field, as shown in **Fig.4b**, is similar to the characteristics of nickel-based high-temperature superconductors.[1,2] The absence of zero resistance under high pressure using paraffin as pressure-transmitting medium is probably due to pressure imhomogenity. Indeed, the effects of pressure-transmitting medium has been discussed.[50] **Fig.4c** shows field dependence of resistivity for Crystal #2 under a pressure of 23.7 GPa using helium as the pressure-transmitting medium. A much sharper transition with ~10 K in width is observed, and a pronounced effect of the suppression of superconductivity by applied magnetic fields was also observed. The use of helium, the best pressure-transmitting medium for hydrostatic pressure, in conjunction with the four-probe method for high-pressure electrical transport measurements on single-crystal samples, ensured the high reliability of the experimental results. Upper critical fields extracted using the normal-state resistance values at 90%, 50%, and 10% of the resistivity near the superconducting transition temperature, represented by open circles, are shown in **Fig.4d**. Using the Ginzburg-Landau model to fit Hc2(T), zero-temperature values of 291.7 T, 126.3 T, and 82.5 T were obtained. We obtained coherence lengths of 1.1 nm, 1.6 nm, and 2.0 nm, which are comparable to those of bulk $La_3Ni_2O_7$[1-2]. The fan-shaped broadening of the superconducting transition under a magnetic field is a typical manifestation of flux creep in high-temperature superconductors. The significant differences in the values of the upper critical field (Hc2) defined by different criteria are a specific reflection of this behavior.

## Conclusion

We succeeded in growing single crystals of bilayer nickelates ($La_{3-x}R_xNi_2O_7$, R= Pr-Er, $x\leq 2.7$) at ambient pressure using flux method, overcoming the previous requirements of high gas pressure for single crystal preparation. The maximum substitution $x$ decreases with the decreasing of the size of rare earth ions; however, the molar ratio of (La+R) to Ni maintains 3:2 and bilayer structure. Among the whole series, we found that the Sm-compound $La_2SmNi_2O_{7-\delta}$ is the best to grow. By suppressing the growth of hybrid R-P phases, the size of single crystals of $La_2SmNi_2O_{7-\delta}$ was increased to 220 μm on the edge. EDS mapping, single crystal X-ray diffraction, NQR and real-space imaging via STEM evidence high quality of our bilayer $La_2SmNi_2O_{7-\delta}$ single crystals. High pressure resistivity measurements revealed that the $T_c^{onset}$ of annealed $La_2SmNi_2O_{7-\delta}$ single crystals reached 91 K at 22 GPa, which is the highest among the known superconducting nickelates. Our results not only provide an easy-to-access strategy to prepare high-quality bilayer nickelate single crystals for unlocking the mystery of high-$T_c$ superconductivity, but also provide a direction for discovering higher-$T_c$ nickelate superconductors.

## Acknowledgements

Work at Shandong University was supported by the National Natural Science Foundation of China (12074219 and 12374457), the TaiShan Scholars Project of Shandong Province (tsqn201909031), and the QiLu Young Scholars Program of Shandong University. This work was supported by the National Key Research and Development Projects of China (Grants No. 2023YFA1406103, No. 2024YFA1611302, No. 2024YFA1409200 and No. 2022YFA1403402), the National Natural Science Foundation of China (Grants No. 12374142, No. 12304170 and No. U23A6003). Q.S.Z. and D.P. acknowledge the support from Shanghai Key Laboratory of Material Frontiers Research in Extreme Environments, China (No. 22dz2260800), the Shanghai Science and Technology Committee, China (No. 22JC1410300). A portion of this work was carried out at the Synergetic Extreme Condition User Facility (SECUF, https://cstr.cn/31123.02.SECUF). J.Z. thanks Dr. Yu-Sheng Chen and Dr. Tieyan Chang from The University of Chicago for stimulating discussions.

## Author contributions

J.Z. conceived the project; F.L. grew single crystals, performed the powder and single-crystal X-ray diffraction experiments, carried out SEM, EDS, magnetic susceptibility, and transport measurements at ambient pressure with the help of C.L, L.W, Jian Z. and J.Z.; D.P. performed the resistance measurements using helium gas as the pressure-transmitting medium under pressure with the help of Q.S.Z.; L.M., J.C. and Z.C. performed high pressure measurements using solid or liquid as pressure-transmitting medium; N.G. and Q.Z. carried out STEM measurements; J.D., J.L., J.Y., and R.Z. performed NQR measurements; F.L., D.P., J.D., L.M., L.W., J.C., Q.Z., R.Z., X.T., and J.Z. discussed and analyzed data; F.L., D.P. and J.Z. wrote the draft with contributions from all coauthors.

## Competing interests

The authors declare no conflict of interest.

**Methods**

**Single crystal growth.** All crystal growth were carried out at ambient pressure. $La_2O_3$ (Sigma-Aldrich, 99.99%) was baked at 600 °C for 5 h before use. Rare Earth Oxide, and NiO (Alfa Aesar, 99.99%) powders were weighed, mixed and ground, and then placed in an $Al_2O_3$ crucible. The mixture was mixed with anhydrous $K_2CO_3$ powders which were used as a flux ($La_{3-x}R_xNi_2O_7$:$K_2CO_3$=1:15, mass ratio). The crucible was covered with a lid in order to minimize the evaporation of $K_2CO_3$. Loading anhydrous $K_2CO_3$ was performed in glove box. Crystal growth was achieved via flux evaporation in a period of 72 h at a temperature of 1000-1050 °C, followed by furnace cooling to room temperature.

**Powder X-ray diffraction (PXRD). A** Bruker AXS D2 Phaser X-ray powder diffractometer was used to check phase purity. Data were collected at room temperature using Cu-$K_\alpha$ radiation ($\lambda$ = 1.5418 Å) in the 2θ range of 20-90° with a scan step size of 0.02° and a scan time of 2 s per step. TOPAS 6 was used for Rietveld refinement. Refined parameters include background (chebychev function, order 13), sample displacement, lattice parameters and strain.

**Single-Crystal Structure Determination.** Single crystal X-ray diffraction data were collected on a Bruker AXS D8 Venture (Mo-$K_\alpha$ radiation, $\lambda$ = 0.71073 Å) diffractometer at 296 K. A single crystal of $La_3Ni_2O_7$ with dimensions of 0.043×0.032×0.022 $mm^3$ and a single crystal of $La_2SmNi_2O_7$ with 0.025×0.033×0.013 $mm^3$ were used. Indexing was performed using Bruker APEX4 software.[51] Data integration and cell refinement were performed using SAINT, and multi-scan absorption corrections were applied using the SADABS program.[51] The structure was solved by direct methods and refined with full matrix least-squares methods on $F^2$. All atoms were modeled using anisotropic ADPs, and the refinements converged for $I > 2\sigma(I)$, where $I$ is the intensity of reflections and σ(I) is standard deviation. Calculations were performed using SHELXTL[51] and Olex2.[52] Further details of the crystal structure investigations may be obtained from the joint CCDC/FIZ Karlsruhe online deposition service by quoting the deposition number 2315725 and 2418256.

**Nuclear quadrupole resonance (NQR).** NQR measurements were conducted using a phase-coherent pulsed NQR spectrometer. $^{139}$La-NQR spectra were obtained by sweeping the frequency point by point, and integrating spin-echo intensity. The quantity of the single crystal sample used for NQR measurement is about 100 mg.

**Scanning transmission electron microscopy (STEM).** $La_3Ni_2O_7$ and $La_2SmNi_2O_7$ single crystals were crushed in ethanol, and drops of the suspensions were deposited on lacey carbon-coated copper grids and dried in air for STEM observations. High-angle annular dark-field (HAADF)-STEM images were obtained at an accelerating voltage of 300 kV on an aberration-corrected transmission electron microscope (Spectra 300, Thermo Fisher Scientific), equipped with a field-emission electron gun. The probe convergence semi-angle and inner collection semi-angle are 25.0 mrad and 49.0 mrad, respectively.

**Scanning Electron Microscopy (SEM).** The morphology of the as-grown crystals was examined using a scanning electron microscope. The scanning electron microscope images were obtained by GeminiSEM-300 microscope incident electron of 15.0 kV.

**Energy Dispersive Spectrometer (EDS).** The X-ray spectrometer Bruker Quantax XFlash6-100 was used for qualitative and quantitative analysis of the as-grown crystals.

**High pressure measurements.** High-pressure resistance measurements on $La_2SmNi_2O_7$ single crystals under pressures up to 24 GPa were carried out with a BeCu-type diamond pressure cell (DAC). The sample was loaded into a pre-indented gasket hole filled with helium as the pressure-transmitting medium in between a pair of diamond anvils with a 400 μm culet. Four gold leads were manually put on the sample surface and the electrical contact maintained by mechanical compression. The van der Pauw four-probe method and the standard four-probe technique were both utilized for resistivity measurements under high pressures. A piece of ruby ball placed near the sample in the DAC is used as the pressure calibrant and the pressure is determined by monitoring the position of the ruby fluorescence R1 line at room temperature. All experiments were carried out in a physical property measurement system by Quantum Design. The temperature range covered was from 2 K to 310 K, and magnetic fields of up to 7 T were applied.

**Data Availability Statement**

The data that support the findings of this study are available in the Supporting Information of this article.

## References

1. Sun, H. L. *et al.* Signatures of superconductivity near 80 K in a nickelate under high pressure. *Nature* **621**, 493-498 (2023). https://doi.org:10.1038/s41586-023-06408-7
2. Zhang, Y. *et al.* High-temperature superconductivity with zero resistance and strange-metal behaviour in $La_3Ni_2O_{7-\delta}$. *Nature Physics* (2024). https://doi.org:10.1038/s41567-024-02515-y
3. Wen, J. *et al.* Probing the Meissner effect in pressurized bilayer nickelate superconductors using diamond quantum sensors. *arXiv: 2410.10275v1* (2024).
4. Keimer, B., Kivelson, S. A., Norman, M. R., Uchida, S. & Zaanen, J. From quantum matter to high-temperature superconductivity in copper oxides. *Nature* **518**, 179-186 (2015). https://doi.org:10.1038/nature14165
5. Pickett, W. E. Colloquium: Room temperature superconductivity: The roles of theory and materials design. *Rev. Mod. Phys.* **95**, 021001 (2023). https://doi.org:10.1103/RevModPhys.95.021001
6. Sanders, S. *125 Questions: Exploration and Discovery*. (Science, 2021).
7. Zhang, J. & Tao, X. Review on quasi-2D square planar nickelates. *CrystEngComm* **23**, 3249-3264 (2021). https://doi.org:10.1039/d0ce01880e
8. Anisimov, V. I., Bukhvalov, D. & Rice, T. M. Electronic structure of possible nickelate analogs to the cuprates. *Physical Review B* **59**, 7901-7906 (1999). https://doi.org:10.1103/PhysRevB.59.7901
9. Zhang, J. *et al.* Large orbital polarization in a metallic square-planar nickelate. *Nature Physics* **13**, 864-869 (2017). https://doi.org:10.1038/nphys4149
10. Li, D. *et al.* Superconductivity in an infinite-layer nickelate. *Nature* **572**, 624-627 (2019). https://doi.org:10.1038/s41586-019-1496-5
11. Botana, A. S., Lee, K.-W., Norman, M. R., Pardo, V. & Pickett, W. E. Low Valence Nickelates: Launching the Nickel Age of Superconductivity. *Frontiers in Physics* **9** (2022). https://doi.org:10.3389/fphy.2021.813532
12. Ji, Y., Liu, J., Li, L. & Liao, Z. Superconductivity in infinite layer nickelates. *J. Appl. Phys.* **130** (2021). https://doi.org:10.1063/5.0056328
13. Wang, B. Y., Lee, K. & Goodge, B. H. Experimental Progress in Superconducting Nickelates. *Annual Review of Condensed Matter Physics* **15**, 305-324 (2024). https://doi.org:https://doi.org/10.1146/annurev-conmatphys-032922-093307
14. Zhou, X. *et al.* Experimental progress on the emergent infinite-layer Ni-based superconductors. *Mater. Today* **55**, 170-185 (2022). https://doi.org:https://doi.org/10.1016/j.mattod.2022.02.016
15. Nomura, Y. & Arita, R. Superconductivity in infinite-layer nickelates. *Rep. Prog. Phys.* **85** (2021).
16. Gu, Q. & Wen, H.-H. Superconductivity in nickel-based 112 systems. *The Innovation* **3** (2021).
17. Chow, S. L. E. & Ariando, A. Nickel Age of High-Temperature Superconductivity. *Advanced Materials Interfaces* **n/a**, 2400717 https://doi.org:https://doi.org/10.1002/admi.202400717
18. Wang, B. X. *et al.* Synthesis and characterization of bulk $Nd_{1-x}Sr_xNiO_2$ and $Nd_{1-x}Sr_xNiO_3$. *Phys Rev Mater* **4** (2020). https://doi.org:10.1103/PhysRevMaterials.4.084409
19. Li, Q. *et al.* Absence of superconductivity in bulk $Nd_{1-x}Sr_xNiO_2$. *Communications Materials* **1** (2020). https://doi.org:10.1038/s43246-020-0018-1
20. Puphal, P. *et al.* Topotactic transformation of single crystals: From perovskite to infinite-layer nickelates. *Sci Adv* **7**, eabl8091 (2021). https://doi.org:10.1126/sciadv.abl8091
21. Chow, S. L. E., Luo, Z. & Ariando, A. High-temperature Superconducting Oxide without Copper at Ambient Pressure. *arXiv: 410.00144v1* (2024).
22. Wang, M., Wen, H.-H., Wu, T., Yao, D.-X. & Xiang, T. Normal and Superconducting Properties of $La_3Ni_2O_7$. *Chinese Physics Letters* **41**, 077402 (2024). https://doi.org:10.1088/0256-307X/41/7/077402
23. Wang, Y., Chen, Z., Zhang, Y. & Jiang, K. The Mottness and the Anderson localization in bilayer nickelate $La_3Ni_2O_{7-\delta}$. *arXiv:2501.08536v1* (2025).
24. Yidi Liu *et al.* Superconductivity and normal-state transport in compressively strained $La_2PrNi_2O_7$ thin films.

*arXiv:2501.08022v1* (2025).

25 Li, P. *et al.* Photoemission evidence for multi-orbital hole-doping in superconducting $La_{2.85}Pr_{0.15}Ni_2O_7$/$SrLaAlO_4$ interfaces. *arXiv:2501.09255v1* (2025).

26 Ko, E. K. *et al.* Signatures of ambient pressure superconductivity in thin film $La_3Ni_2O_7$. *Nature* (2024). https://doi.org:10.1038/s41586-024-08525-3

27 Chen, X. *et al.* Non-bulk Superconductivity in $Pr_4Ni_3O_{10}$ Single Crystals Under Pressure. *arXiV:2410.10666v1* (2025).

28 Shi, M. *et al.* Absence of superconductivity and density-wave transition in ambient-pressure tetragonal $La_4Ni_3O_{10}$. *arXiv：2501.12647v1* (2024).

29 Bhatt, L. *et al.* Resolving Structural Origins for Superconductivity in Strain-Engineered $La_3Ni_2O_7$ Thin Films. *arXiv:2501.08204* (2025).

30 Zhou, G. *et al.* Ambient-pressure superconductivity onset above 40 K in bilayer nickelate ultrathin films. *arXiv：2412.16622v1* (2024).

31 Wang, N. *et al.* Bulk high-temperature superconductivity in pressurized tetragonal $La_2PrNi_2O_7$. *Nature* **634**, 579-584 (2024). https://doi.org:10.1038/s41586-024-07996-8

32 Li, J. *et al.* Pressure-driven right-triangle shape superconductivity in bilayer nickelate $La_3Ni_2O_7$. *arXiv: 2404.11369v2* (2024).

33 Zhang, M. *et al.* Effects of pressure and doping on Ruddlesden-Popper phases $La_{n+1}Ni_nO_{3n+1}$. *Journal of Materials Science & Technology* **185**, 147-154 (2024). https://doi.org:https://doi.org/10.1016/j.jmst.2023.11.011

34 Chu, C. W., Deng, L. Z. & Lv, B. Hole-doped cuprate high temperature superconductors. *Physica C: Superconductivity and its Applications* **514**, 290-313 (2015). https://doi.org:https://doi.org/10.1016/j.physc.2015.02.047

35 Zhu, Y. *et al.* Superconductivity in pressurized trilayer $La_4Ni_3O_{10-\delta}$ single crystals. *Nature* **631**, 531-536 (2024). https://doi.org:10.1038/s41586-024-07553-3

36 Pei, C. *et al.* Pressure-Induced Superconductivity in $Pr_4Ni_3O_{10}$ Single Crystals. *arXiv:2411.08677v1* (2024).

37 Wang, G. *et al.* Pressure-induced superconductivity in polycrystalline $La_3Ni_2O_{7-\delta}$. *Physical Review X* **14**, 011040 (2024). https://doi.org:10.1103/PhysRevX.14.011040

38 Ren, Z.-A. *et al.* Superconductivity at 55 K in Iron-Based F-Doped Layered Quaternary Compound $Sm[O_{1-x}F_x]$ FeAs. *Chinese Physics Letters* **25**, 2215 (2008). https://doi.org:10.1088/0256-307X/25/6/080

39 Zhang, Y., Lin, L.-F., Moreo, A., Maier, T. A. & Dagotto, E. Trends in electronic structures and $s\pm$-wave pairing for the rare-earth series in bilayer nickelate superconductor $R_3Ni_2O_7$. *Physical Review B* **108**, 165141 (2023). https://doi.org:10.1103/PhysRevB.108.165141

40 Pan, Z., Lu, C., Yang, F. & Wu, C. Effect of Rare-Earth Element Substitution in Superconducting $R_3Ni_2O_7$ under Pressure. *Chinese Physics Letters* **41**, 087401 (2024). https://doi.org:10.1088/0256-307X/41/8/087401

41 Liu, Z. *et al.* Evidence for charge and spin density waves in single crystals of $La_3Ni_2O_7$ and $La_3Ni_2O_6$. *Science China Physics, Mechanics & Astronomy* **66**, 217411 (2023). https://doi.org:10.1007/s11433-022-1962-4

42 Zhang, J. *et al.* High oxygen pressure floating zone growth and crystal structure of the metallic nickelates $R_4Ni_3O_{10}$(R=La,Pr). *Physical Review Materials* **4**, 083402 (2020). https://doi.org:10.1103/PhysRevMaterials.4.083402

43 Dong, Z. *et al.* Visualization of oxygen vacancies and self-doped ligand holes in $La_3Ni_2O_{7-\delta}$. *Nature* (2024). https://doi.org:10.1038/s41586-024-07482-1

44 Chen, X. *et al.* Polymorphism in the Ruddlesden–Popper Nickelate $La_3Ni_2O_7$: Discovery of a Hidden Phase with Distinctive Layer Stacking. *Journal of the American Chemical Society* **146**, 3640-3645 (2024). https://doi.org:10.1021/jacs.3c14052

45 Puphal, P. *et al.* Unconventional Crystal Structure of the High-Pressure Superconductor $La_3Ni_2O_7$. *Phys. Rev. Lett.* **133**, 146002 (2024). https://doi.org:10.1103/PhysRevLett.133.146002

46 Wang, H., Chen, L., Rutherford, A., Zhou, H. & Xie, W. Long-Range Structural Order in a Hidden Phase of Ruddlesden–Popper Bilayer Nickelate $La_3Ni_2O_7$. *Inorganic Chemistry* **63**, 5020-5026 (2024). https://doi.org:10.1021/acs.inorgchem.3c04474

47 Li, F. *et al.* Flux Growth of Trilayer $La_4Ni_3O_{10}$ Single Crystals at Ambient Pressure. *Crystal Growth & Design* **24**, 347-354 (2024). https://doi.org:10.1021/acs.cgd.3c01049

48 Li, F. *et al.* Design and synthesis of three-dimensional hybrid Ruddlesden-Popper nickelate single crystals. *Physical Review Materials* **8**, 053401 (2024). https://doi.org:10.1103/PhysRevMaterials.8.053401

49 Luo, J. *et al.* Microscopic evidence of charge- and spin-density waves in $La_3Ni_2O_{7-\delta}$ revealed by $^{139}$La-NQR. *arXiv：2501.11248v1* (2024).

50 Hou, J. *et al.* Emergence of High-Temperature Superconducting Phase in Pressurized $La_3Ni_2O_7$ Crystals. *Chinese Physics Letters* **40**, 117302 (2023). https://doi.org:10.1088/0256-307X/40/11/117302

51 Computer code APEX5 (Bruker Analytical X-ray Instruments, Inc. Madison, Wisconsin, USA., 2023).

52 Dolomanov, O. V., Bourhis, L. J., Gildea, R. J., Howard, J. A. K. & Puschmann, H. OLEX2: a complete structure solution, refinement and analysis program. *J. Appl. Crystallogr.* **42**, 339-341 (2009). https://doi.org:https://doi.org/10.1107/S0021889808042726

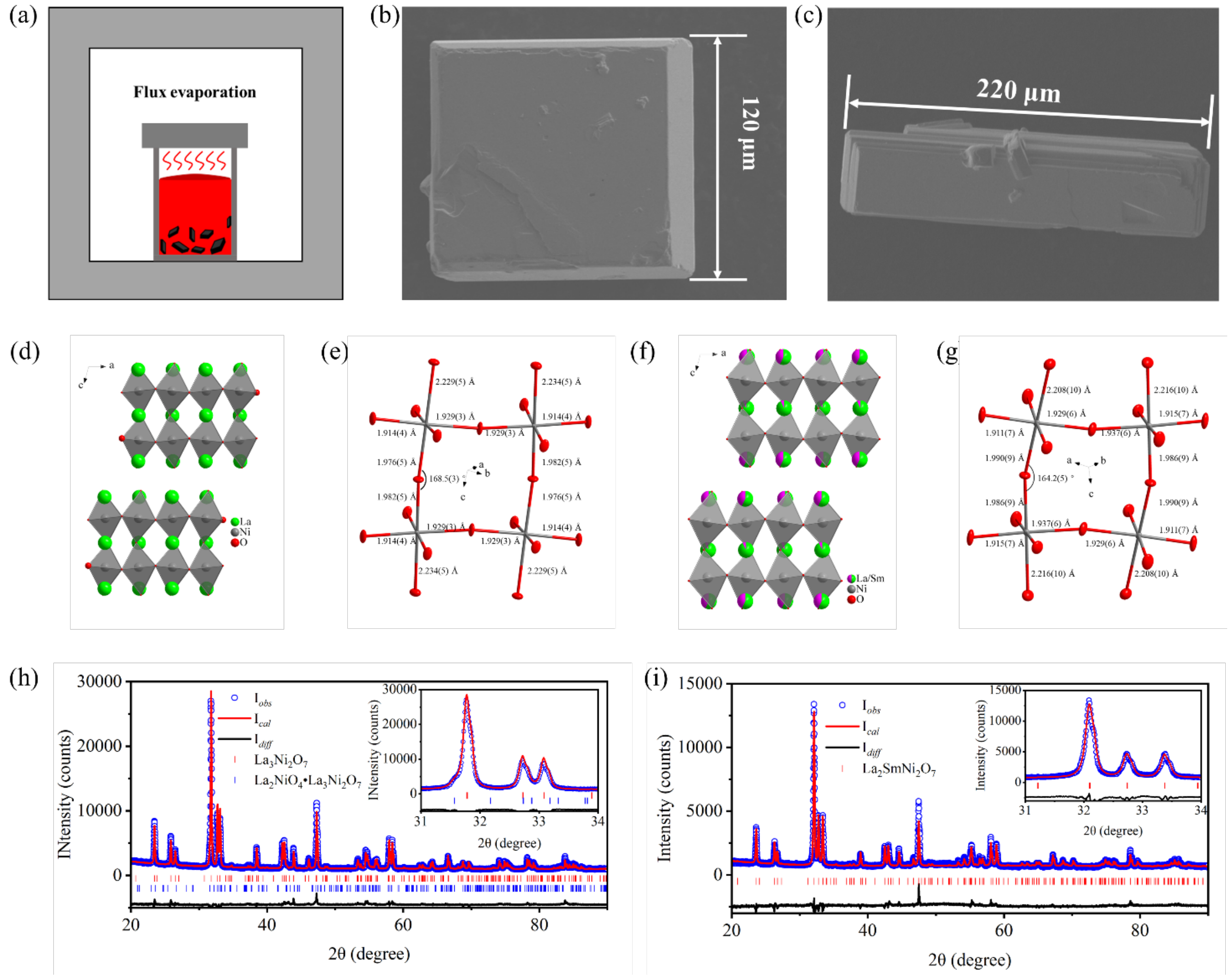

**Figure 1**. Ambient flux growth of bilayer nickelate single crystals. (a) A scheme of crystal growth using flux method via evaporation at ambient pressure. (b) A scanning electron microscopy (SEM) image of a typical $La_3Ni_2O_{7-\delta}$ single crystal. (c) A SEM image of a typical $La_2SmNi_2O_{7-\delta}$ single crystal. (d) Crystal structure of $La_3Ni_2O_7$ in the polyhedral mode obtained from X-ray single crystal diffraction. (e) Ellipsoid drawings of the NiO6 octahedra with bond distances and bond angles of $La_3Ni_2O_{7-\delta}$. (f) Crystal structure of $La_2SmNi_2O_{7-\delta}$ obtained from X-ray single crystal diffraction. (g) Ellipsoid drawings of the NiO6 octahedra with bond distances and bond angles in the bilayer NiO2 of $La_2SmNi_2O_{7-\delta}$. (h) X-ray powder diffraction data collected at room temperature and Rietveld refinement of pulverized $La_3Ni_2O_{7-\delta}$ single crystals. (i) X-ray powder diffraction data collected at room temperature and Rietveld refinement of pulverized $La_{3-x}Sm_xNi_2O_{7-\delta}$ single crystals.

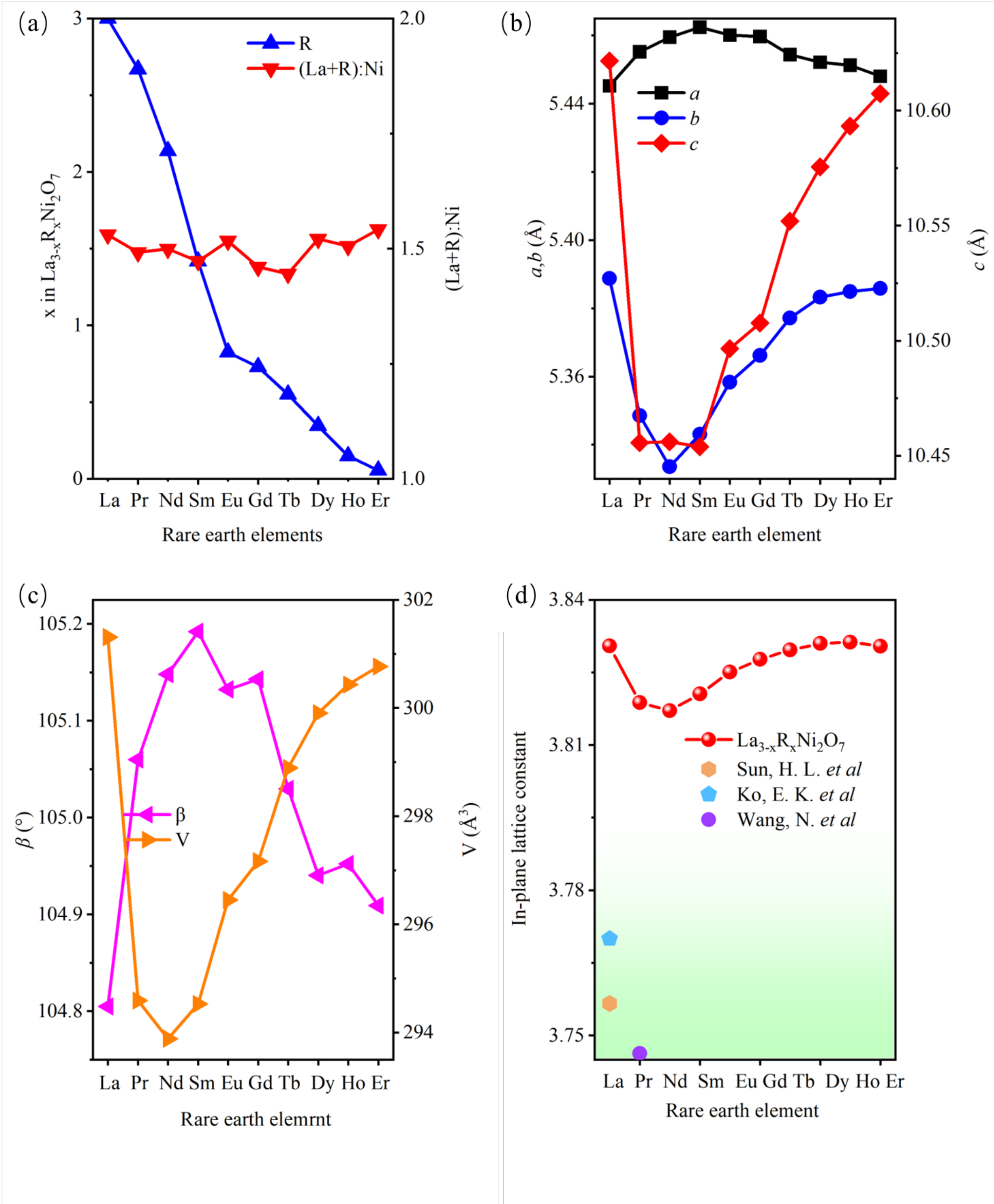

**Figure 2**. Lattice parameters extracted from LeBail fit on in-house X-ray powder diffraction data of pulverized $La_{3-x}R_xNi_2O_{7-\delta}$ (R=La-Er) single crystals. (a) The maximum substitution and molar ratio of (La+R):Ni obtained from EDS for different rare earth elements in as-grown $La_{3-x}R_xNi_2O_{7-\delta}$ (R=La-Er) single crystals. (b,c) The lattice parameters extracted from LeBail fit on in-house X-ray powder diffraction data for maximum substitution of as-grown $La_{3-x}R_xNi_2O_{7-\delta}$ (R=Pr-Er). (d) The variation of averaged in-plane lattice constant of $La_{3-x}R_xNi_2O_{7-\delta}$ (R=La-Er) calculated by $a_{average} = (a^2 + b^2)^{1/2}/2$ with those of superconducting states for comparison. Note Sun et al. for Ref.[1] Ko et al. for Ref.[26]and Wang et al. for Ref.[31]

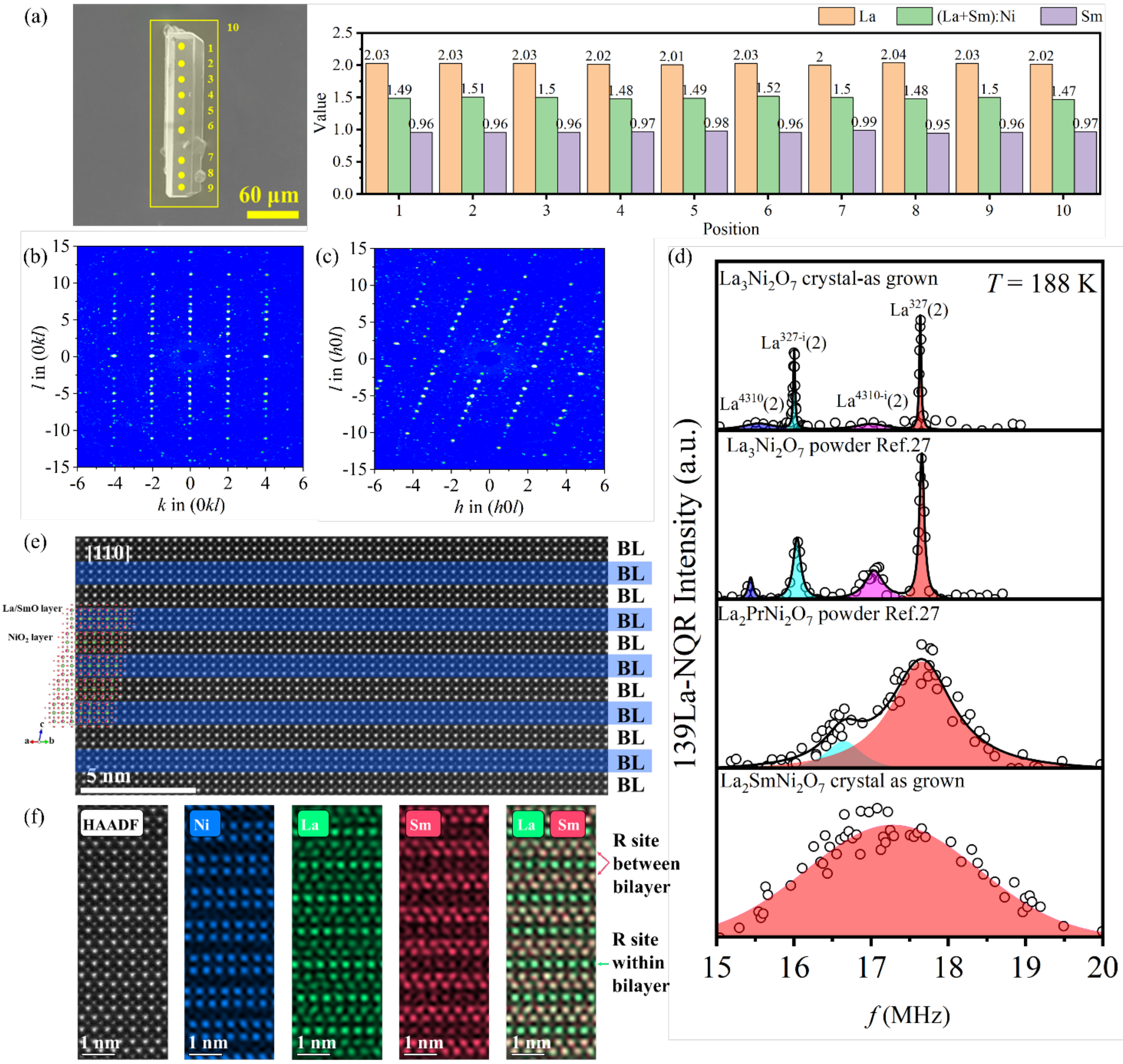

**Figure 3**. Crystal quality of $La_2SmNi_2O_{7-\delta}$ single crystals. (a) EDS mapping of a typical $La_2SmNi_2O_7$ single crystal. (b) Reconstructed (0*kl*) plane from in-house X-ray single crystal diffraction data of $La_2SmNi_2O_{7-\delta}$ single crystal collected at 296(2) K. (c) Reconstructed (*h*0*l*) plane from in-house X-ray single crystal diffraction data of $La_2SmNi_2O_7$ single crystal collected at 296(2) K. (d) $^{139}$La (2) NQR spectra corresponding to the ±5/2↔±7/2 transition in $La_3Ni_2O_{7-\delta}$ and $La_2SmNi_2O_{7-\delta}$ crystals at 188 K. The solid lines represent fits using Lorentz and Gaussian functions for $La_3Ni_2O_{7-\delta}$ and $La_2SmNi_2O_{7-\delta}$, respectively. (e) A typical atomic-scale HAADF-STEM image in the projection of [110] with overlaid crystal structure model of $La_2SmNi_2O_{7-\delta}$ single crystal. (f) EDS maps from STEM for La, Sm and Ni and mixed color map of them.

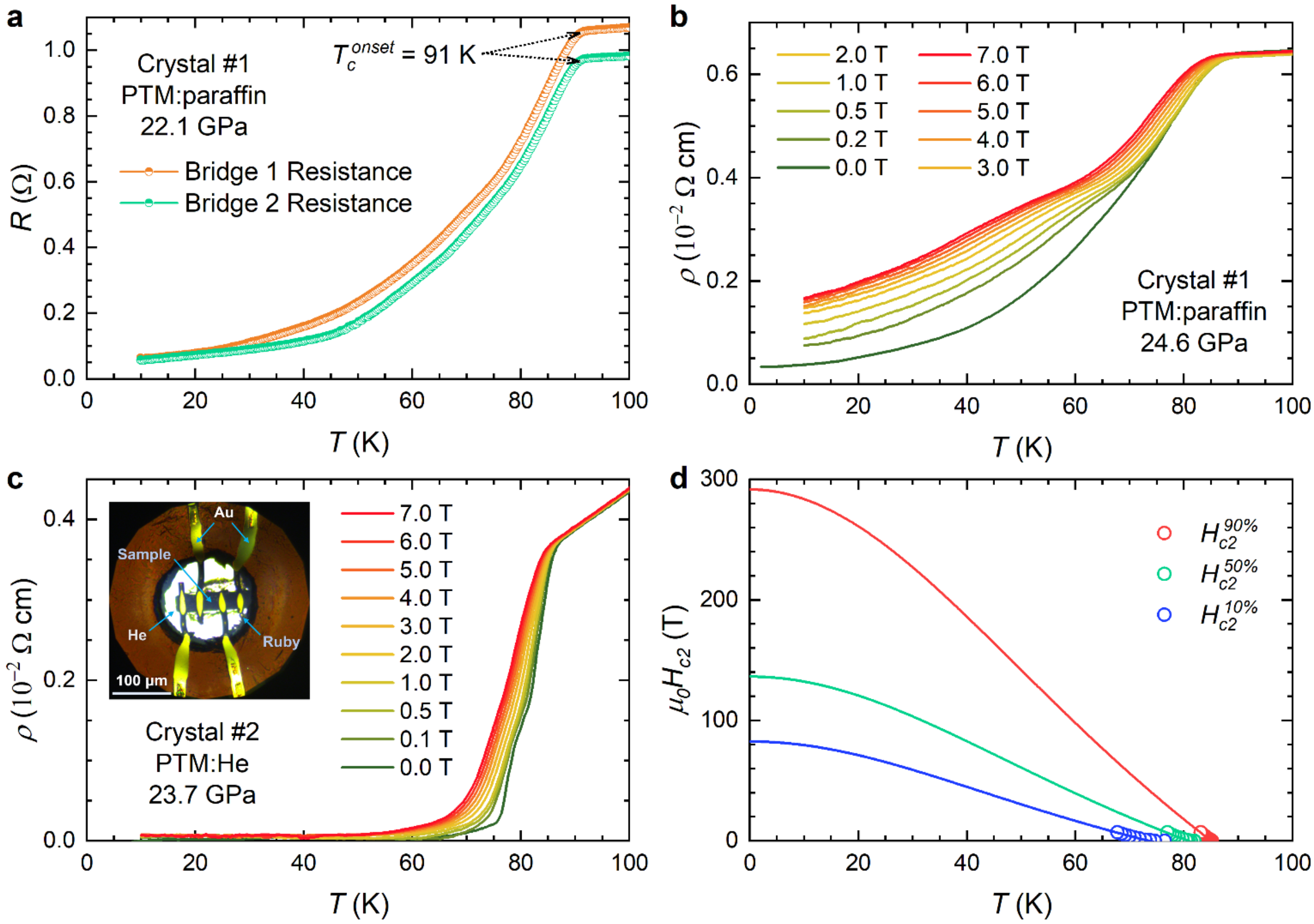

**Figure 4**. Pressure-induced high-$T_c$ superconductivity in $La_2SmNi_2O_{7-\delta}$. (a) Temperature-dependent resistance of Crystal #1 under pressure in a paraffin-filled diamond anvil cell (DAC). The resistance of Bridge 1 and Bridge 2 represents the electrical resistance measured along two orthogonal directions separated by approximately a 90º angle. (b) Magnetic field effects on the superconducting transition in Crystal #1. (c) Field dependence of resistivity for Crystal #2 at 23.7 GPa, with helium as the pressure-transmitting medium. Inset: a photograph of the electrodes used for high-pressure resistivity measurements. Helium as the pressure-transmitting medium under high pressure maintained the integrity of the single-crystal sample, and direct contact was established between the sample and the gold electrodes. (d) Upper critical fields extracted using the normal-state resistance values at 90%, 50%, and 10% of the resistivity near the superconducting transition temperature, represented by open circles. Solid lines show the results of fitting with a Ginzburg–Landau model.

**Table 1** Summary of crystal growth of bilayer nickelates via different methods

<table>
<tr><th>Materials</th><th>Growth method</th><th>La:R (molar ratio)</th><th>T (ºC)</th><th>Space group</th><th>Main advantage</th><th>Main issues/ comments</th><th>Ref.</th></tr>
<tr><td>$La_3Ni_2O_{7-\delta}$</td><td>Floating zone growth at 10-18 bar $O_2$</td><td>-</td><td>-</td><td><i>Cmcm</i></td><td>grow fast</td><td>High $pO_2$, narrow $pO_2$ range, difficult to obtain pure phase due to the competition of “1313”, intergrowth, expensive furnace not easy to access</td><td>1,2,46,44,45</td></tr>
<tr><td>$La_3Ni_2O_{7-\delta}$</td><td rowspan="16">Flux growth at 0.2 bar $O_2$ using $K_2CO_3$ as a flux</td><td>-</td><td rowspan="16">1000 ~1050</td><td rowspan="16">$P2_1/m$</td><td rowspan="16">Ambient pressure, easy to access</td><td rowspan="8">Competition of hybrid R-P phases</td><td rowspan="16">This work</td></tr>
<tr><td>$La_{2.05}Pr_{0.95}Ni_2O_{7-\delta}$</td><td>2:1</td></tr>
<tr><td>$La_{1.01}Pr_{1.99}Ni_2O_{7-\delta}$</td><td>1:2</td></tr>
<tr><td>$La_{0.45}Pr_{2.55}Ni_2O_{7-\delta}$</td><td>1:5</td></tr>
<tr><td>$La_{0.3}Pr_{2.7}Ni_2O_{7-\delta}$</td><td>1:9</td></tr>
<tr><td>$La_{2.15}Nd_{0.85}Ni_2O_{7-\delta}$</td><td>2:1</td></tr>
<tr><td>$La_{1.05}Nd_{1.95}Ni_2O_{7-\delta}$</td><td>1:2</td></tr>
<tr><td>$La_{0.87}Nd_{2.13}Ni_2O_{7-\delta}$</td><td>7:23</td></tr>
<tr><td><b>$La_{2.03}Sm_{0.97}Ni_2O_{7-\delta}$</b></td><td>2:1</td><td><b>The best to grow</b>.</td></tr>
<tr><td>$La_{1.57}Sm_{1.43}Ni_2O_{7-\delta}$</td><td>1:2</td><td rowspan="7">Competition of hybrid R-P phases</td></tr>
<tr><td>$La_{2.17}Eu_{0.83}Ni_2O_{7-\delta}$</td><td rowspan="6">2:1</td></tr>
<tr><td>$La_{2.27}Gd_{0.73}Ni_2O_{7-\delta}$</td></tr>
<tr><td>$La_{2.45}Tb_{0.55}Ni_2O_{7-\delta}$</td></tr>
<tr><td>$La_{2.65}Dy_{0.35}Ni_2O_{7-\delta}$</td></tr>
<tr><td>$La_{2.85}Ho_{0.15}Ni_2O_{7-\delta}$</td></tr>
<tr><td>$La_{2.95}Er_{0.05}Ni_2O_{7-\delta}$</td></tr>
</table>